
\magnification=1300

\headline{\ifnum\pageno=1 \nopagenumbers
\else \hss\number \pageno \fi}

\overfullrule=0pt

\font\boldgreek=cmmib10
\textfont9=\boldgreek
\mathchardef\mymu="0916 
\footline={\hfil}
\baselineskip=12pt

\def \cc #1 {\kern .7em \hfill #1 \hfill \kern .7em}
\vbox to 1,5truecm{}

\input epsf

\parskip=0.2truecm
\centerline{\bf Strange Baryon Production}
\medskip
 \centerline{\bf in Heavy Ion Collisions}\bigskip

\bigskip \centerline{by}\smallskip
\centerline{{\bf A. Capella, A. Kaidalov}\footnote{*}{Present address~:
ITEP, B.
Cheremushkinskaya ulitsa 25, 117 259 Moscow, Russia}{\bf, A. Kouider Akil,
C. Merino, J. Tran
Thanh Van}}  \medskip
 \centerline{Laboratoire de Physique Th\'eorique et Hautes Energies
\footnote{**}{Laboratoire associ\'e au Centre National de la Recherche
Scientifique - URA D0063}}  \centerline{Universit\'e de Paris XI, b\^atiment
211, 91405
Orsay Cedex, France}
\bigskip\bigskip
\baselineskip=20pt
\noindent
${\bf Abstract}$ \par
The rapidity distribution of $\Lambda$ and $\bar{\Lambda}$ produced in
nucleus-nucleus
collisions at CERN energies is studied in the framework of an independent
string model -
with quark-antiquark as well as diquark-antidiquark pairs in the nucleon
sea. It is shown
that, besides the $\Lambda$-$\bar{\Lambda}$ pair production resulting from
the fragmentation
of sea diquarks, final state interactions of co-moving secondaries $\pi + N
\to K + \Lambda$
and $\pi + \bar{N} \to K + \bar{\Lambda}$ are needed in order to reproduce
the
data. Predictions for $Pb$-$Pb$ collisions are presented.

\vbox to 1cm{}

\noindent LPTHE Orsay 95-41 \par
\noindent June 1995
\vfill\supereject \noindent {\bf 1. Introduction} \medskip
A substantial increase in the ratio of strange over non-strange particles
between
proton-proton and central nucleus-nucleus collisions has been observed
experimentally both at
CERN and AGS [1]. Such an increase was expected as a consequence of quark
gluon plasma
(QGP) formation due to an increase of the strange quark production rate in
the deconfined
phase [2]. However, as in the case of other possible signals of QGP, one has
to check whether
such a phenomenon is typical of a central nucleus-nucleus collision or, on
the
contrary, it is already present in proton-nucleus interactions
\footnote{*}{For instance, in
the case of $J/\psi$ suppression, R. Salmeron, followed by many other
authors, has shown
that the effect could be explained by extrapolation of the corresponding
hadron-nucleus data
[3].}. Unfortunately, the experimental situation is not conclusive. While in
the data from
the NA35 collaboration strangeness is enhanced between $pp$ and $pS$, the
effect is not seen
when taking an average over three different nuclear targets [4]. However,
two other
experimental facts indicate that strangeness enhancement is not only present
in central
nucleus-nucleus collisions. Indeed in $pp$ collisions, the ratio $K/\pi$
increases both with
increasing energy and with increasing multiplicity [5]. Also the ratio
$\bar{\Lambda}/\bar{p}$ in $pp$ collisions increases from 0.26 at $\sqrt{s}
=$ 20 GeV [6] to
0.39 at $\sqrt{s} =$ 1 800 GeV [7]. Furthermore, results from the NA36
collaboration [8] show
a large strangeness enhancement between peripheral and average
nucleus-nucleus collisions.
\par

Actually, it has been emphasized [9-11] that strangeness enhancement is
expected to occur
in independent string models such as the Dual Parton Model (DPM) [9] and the
Quark Gluon
String Model (QGSM) [12]. Indeed, the average number of collisions per
participant nucleon
increases between $pp$ and $pA$ as well as between $pA$ and central $AB$
collisions. The
resulting extra particles are produced in the fragmentation of strings
involving sea
constituents at their ends. Since strange sea quarks are present in the
nucleon sea, the
ratio of strange over non-strange secondaries will also increase. The value
of this
increase depends on the ratio
$$S = {2 (s + \bar{s}) \over u + d + \bar{u} + \bar{d}}$$

\noindent in the nucleon sea. We shall use a value of $S$ in the range 0.4
to 0.5,
consistent both with strangeness suppression in string fragmentation at the
highest
available energies [7] and with conventional parton distributions for DIS
[13]. In order to
explain the observed enhancement of $\Lambda$ and $\bar{\Lambda}$, two
approaches have been
proposed. In DPM, diquark-antidiquark pairs have been introduced in the
nucleon sea - with
the same ratio, relative to quark-antiquark pairs, as in the string breaking
process [10,
11]. The second approach is string fusion [14, 15]. However, both mechanisms
have a common
caveat, namely, they lead to the same increase in \underbar{absolute value}
of strange
baryons and antibaryons - while experimentally the excess of $\Lambda$'s is
much larger than
that of $\bar{\Lambda}$'s.  \par

In the present work we solve this problem by introducing final state
interactions of
co-moving pions and nucleons which increase the number of produced $
\Lambda$'s via the
binary interactions $\pi + N \to K + \Lambda$ ($K^* + \Lambda$, ...). This
is a
fundamental mechanism in hadron gas models [2] and has been incorporated
into some Monte
Carlo codes [12, 15, 16]. Clearly, such an interaction is very efficient in
producing
$\Lambda$'s - since the density of both pions and nucleons is high as
compared to that
of $\Lambda$'s. Actually, it was shown in Ref. [11], that using the
experimentally
measured pion and nucleon densities in central $SS$ collisions, together
with the
cross-section of the above reaction (which is about 1.5 mb at the maximum
near threshold), one
precisely obtains the missing number of $\Lambda$'s. Obviously, the
conjugate reaction $\pi +
\bar{N} \to K + \bar{\Lambda}$ ($K^* + \bar{\Lambda}$) will also produce
some extra
$\bar{\Lambda}$ but the absolute yield is much smaller than the one of $
\Lambda$'s due to the
smallness of the $\bar{N}$ density as compared to that of $N$'s. \par

In this paper we compute, in the framework described above, the rapidity
distributions of
$\Lambda$ and $\bar{\Lambda}$ in $SS$ and $Pb$-$Pb$ collisions. A
substantial increase of the
ratios $\Lambda/\pi^-$ and $\bar{\Lambda}/\pi^-$ (70 $\%$ and 50 $\%$,
respectively) is
predicted between $SS$ and $Pb$-$Pb$. \par

Extra evidence for the necessity of a final state interaction is obtained
from baryon
stopping. The rapidity distribution of proton minus antiproton in central
$SS$ collisions,
computed in independent string models, has a dip at $y^* \sim 0$ which is
much more
pronounced that the one observed experimentally [17] - while the
corresponding distributions
both in $pp$ and in peripheral $SS$ collisions are well reproduced. In other
words, the
stopping power in the model is too small as compared to experiment and
rescattering of the
produced baryon is required in order to increase it [12] [15]. \par

The plan of the paper is as follows~: in Section 2 we give the general
formulae of DPM
for nucleus-nucleus collisions in the presence of diquark-antidiquark pairs
in the nucleon
sea and compute the rapidity distributions of $\Lambda$ and $\bar{\Lambda}$
in $pp$ and
$SS$ collisions without final state interactions. In Section~3, we introduce
the final state
interactions and give the modified $\Lambda$ and $\bar{\Lambda}$ rapidity
distributions in $SS$ collisions as well as the model predictions for
$Pb$-$Pb$ collisions.
The rapidity distribution of pions and protons in $pp$, $SS$ and $Pb$-$Pb$
collisions are
also given. A list of average multiplicities for pions, kaons, baryons and
antibaryons
without and with final state interactions is given in Table~1. Section~4
contains some final
remarks and conclusions. In Appendix 1 we give the momentum distribution
functions and
fragmentation functions used in this work, as well as the formulae for the
hadronic spectra of
the individual strings.

\par \medskip
\noindent {\bf 2. The Model} \medskip
In DPM and QGSM, the dominant contribution to multiparticle production in
$pp$ collisions is
a two string production - the two strings being stretched between the
valence diquark of a
proton and a valence quark of the other proton, and vice-versa. In fact, in
this paper we
shall restrict ourselves to this component (see Section 4  for a discussion
on this
approximation). Therefore we have exactly two strings per inelastic
collision. When the
number of inelastic collisions experienced by a nucleon in a nuclear
collision is larger
than one, the extra strings (two per each extra collision) have sea quark
constituents at
the end corresponding to the multiply wounded nucleon. As explained in the
Introduction,
since strange sea quarks are present in the nucleon sea, the ratio of
strange over
non-strange particles will increase. One can object, however, that if the
fraction $S$ of
strange quarks (defined in the Introduction) is the same as in the string
breaking process,
then the strings with sea quarks at their ends will produce the same strange
over non-strange
ratio as in the string breaking process - with no net strangeness
enhancement. While true at
asymptotic energies, this is not the case at present ones. For instance, let
us consider the
$K^-/\pi^-$ ratio. In $NN$ collisions at 200 GeV/c, the experimental value
of this ratio is
about 0.05. Since the values of $S$ we consider are much larger, it is clear
that the ratio
$K^-/\pi^-$ will increase. Another example is the ratio $\bar{\Lambda}/
\bar{p}$ which, as
discussed in the Introduction, reaches a value close to 0.4 at Fermilab [7].
In the following
we shall use the value $S = 0.5$. However, taking $S = 0.4$ changes very
little our results
[11]. \par

The general formulae for the single particle inclusive rapidity distribution
in nucleus-nucleus collisions is the following [9]~:
$${dN^{AB} \over dy} = {1 \over \sigma_{AB}} \sum_{n_A, n_B, n}
\sigma_{n_A,n_B,n} \left \{
\theta (n_B - n_A) \left [ n_A\left ( N_{\mu_A, \mu_B}^{qq_v^A-q_v^B}(y) +
N_{\mu_A,
\mu_B}^{q_v^A-qq_v^B}(y) \right ) + \right .\right .$$
$$\left . (n_B - n_A) \left ( N_{\mu_A, \mu_B}^{\bar{q}_s^A-q_v^B}(y) + N_{
\mu_A,
\mu_B}^{q_s^A-qq_v^B} \right ) + (n - n_B) \left ( N_{\mu_A, \mu_B}^{q_s^A-
\bar{q}_s^B}(y) +
N_{\mu_A, \mu_B}^{\bar{q}_s^A-q_s^B}(y) \right ) \right ]$$ $$\left . + sym
\ (n_A
\leftrightarrow n_B) \right \} \ \ \ .\eqno(2.1)$$

\noindent Here $\sigma_{n_A,n_b,n}^{AB}$ is the cross-section for $n$
inelastic
nucleon-nucleon collisions involving $n_A$ and $n_B$ participating nucleons,
$\sigma_{AB} =
\sum\limits_{n_A,n_B,n} \sigma_{n,n_A,n_B}$ is the inelastic nucleus-nucleus
cross-section,
and $\mu_A = n/n_A$, $\mu_B = n/n_B$. $N(y)$ are the inclusive spectra of
each individual
string. Note that the total number of strings in Eq. (2.1) is $2n$ (two for
each
nucleon-nucleon collision). \par

In the following we restrict the discussion to the case $A = B$. Eq. (2.1)
can be written in a
simple, albeit approximate, form $${dN^{A_PA_T} \over dy} = \bar{n}_A \left
( N_{\bar{\mu}_A,
\bar{\mu}_A}^{qq_v^P-q_v^T}(y) + N_{\bar{\mu}_A, \bar{
\mu}_A}^{q_v^P-qq_v^T}(y) \right
)$$
$$+ (\bar{n} - \bar{n}_A) \left ( N_{\bar{\mu}_A, \bar{\mu}_A}^{q_s^P-
\bar{q}_s^T}(y) +
N_{\bar{\mu}_A, \bar{\mu}_A}^{\bar{q}_s^P-q_s^T}(y) \right ) \ \ \ .
\eqno(2.2)$$

Eq. (2.2) is obtained by neglecting the dependence of the string densities
$N$ on the
$\mu$'s (which are replaced by their average values $\bar{\mu}$). The
summations in (2.2)
are then trivial - the numbers $n_A$ and $n$ being replaced by their average
values
$\bar{n}_A$ and $\bar{n}$. The approximate Eq. (2.2) is quite good at
mid-rapidities. At
large $y$ it becomes worse. However, in the rapidity region where the baryon
densities are
sizeable ($y^* \leq 2.5$), the errors involved are 5 to 10 $\%$. \par

In Eqs. (2.1) and (2.2) only $q_s$-$\bar{q}_s$ pairs in the nucleon sea are
considered. As
discussed in the Introduction, we now assume that $qq_s$-$\bar{q}\bar{q}_s$
pairs are also
present with relative fraction $\alpha$. The modification in Eqs. (2.1) and
(2.2) are
straightforward. Eq. (2.2) is changed into~:

$${dN^{A_PA_T} \over dy} = \bar{n}_A \left ( N_{\bar{\mu}_A,
\bar{\mu}_A}^{qq_v^P-q_v^T} (y) + N_{\bar{\mu}_A, \bar{
\mu}_A}^{q_v^P-qq_v^T} (y) \right ) +
(\bar{n} - \bar{n}_A) \left [ (1 - 2 \alpha ) \left (
N_{\bar{\mu}_A,\bar{\mu}_A}^{q_s^P-\bar{q}_s^T}(y) \right . \right .$$
$$\left . \left . + N_{\bar{\mu}_A ,
\bar{\mu}_A}^{\bar{q}_s^P-q_s^T}(y) \right ) + \alpha \left (
N_{\bar{\mu}_A, \bar{\mu}_A}^{qq_s^P-q_s^T}(y) + N_{\bar{\mu}_A,
\bar{\mu}_A}^{\overline{qq}_s^P-\bar{q}_s^T}(y) + N_{\bar{\mu}_A,
\bar{\mu}_A}^{q_s^P-qq_s^T}(y) + N_{\bar{\mu}_A,
\bar{\mu}_A}^{\bar{q}_s^P-\overline{qq}_s^T}(y) \right ) \right ] \ .
\eqno(2.3)$$

\noindent In the following, we compute the rapidity distributions of $
\Lambda$ and
$\bar{\Lambda}$ in nucleus-nucleus collisions using Eq. (2.3). The rapidity
distributions
of the individual strings are obtained by convoluting momentum distribution
functions and
fragmentation functions. All relevant formulae (including absolute
normalization) are given
in Appendix 1. \par

In numerical calculations we use the value $\alpha = 0.1$ taken from JETSET
[18]. (A somewhat
smaller value, $\alpha = 0.07$, is obtained in FRITIOF [16] and in BAMJET
[19]). The values of
$\bar{n}_A$ and $\bar{n}$ are obtained in the optical approximation of the
Glauber model, with
Saxon-Woods profiles, for impact parameter $|b| \leq$ 1 fm. We get in $SS$
collisions~:
$\bar{n}_A = 31$, $\bar{n} = 55$ and in $Pb$-$Pb$ collisions $\bar{n}_A =
192$, $\bar{n} =
676$. \par

Before giving the numerical results, let us make some remarks concerning
Eq.~(2.3).
Its first two terms are the same as in $NN$ collisions - except for the
value of
$\bar{\mu}_A$ - which is 1 for $NN$, 2 for central $SS$ and 4.5 for central
$Pb$-$Pb$.
The third and fourth terms give negligeably small contributions to
baryon-antibaryon
production due to the large, $(m_{\Lambda} + m_{\bar{\Lambda}})^2$,
thresholds in
the short $q_s$-$\bar{q}_s$ strings. The remaining terms contain the
contribution of sea
diquarks and produce $\Lambda$ and $\bar{\Lambda}$ by pairs. Detailed
formulae for the
rapidity distributions of baryons and antibaryons resulting from the various
terms of Eq.
(2.3) are given in Appendix 1. \par

In Fig. 1 we give the Feynman $x$ distributions of $\Lambda$ and $\bar{
\Lambda}$ in $pp$
collisions. They are obtained from the first two terms of Eq. (2.3) (i.e.
from the sum of
the contributions of the two $qq$-$q$ valence-valence strings) with $\bar{
\mu}_A = 1$. As
shown previously in Ref. [20], the agreement with available data is quite
good. Note that the
shape of the fragmentation functions is determined from Regge intercepts
(see Ref. [20] and Appendix 1). \par

The corresponding rapidity distributions in central $SS$ collisions are
given in Figs.~2 and
3. We see that the yields of both $\Lambda$ and $\bar{\Lambda}$ are smaller
than the data.
The $\Lambda$ yield, in particular, is considerably smaller than the
experimental one. Therefore, we need a mechanism to produce extra $
\Lambda$'s. As already
mentioned in the Introduction this can be achieved by a final state
interaction of
co-moving secondaries $\pi + N \to K + \Lambda$ ($K^* + \Lambda$, ...). A
smaller number of
$\bar{\Lambda}$ will also be produced via $\pi + \bar{N} \to K + \bar{
\Lambda}$ ($K^* +
\bar{\Lambda}$, ...). In the next section we discuss in detail these final
state
interactions. \par  \medskip

\noindent {\bf 3. The Final State Interaction of Secondaries} \medskip
In independent string models, it is assumed that secondaries produced in
different strings
are independent from each other. Introducing a final state interaction of
secondaries
represents a departure from the independent string picture. In fact, it
corresponds to the
simplest form of string interaction. It should be stressed that it is not
known how to treat
quantitatively such an interaction. Here we adopt the formalism introduced
in Ref. [21].
Note that, at moderate energies, such a treatment of string interaction can
be justified in
the framework of Reggeon Field Theory [22]. \par

Let us consider the number of $\Lambda$'s produced by unit of space time
volume $d^4x$
resulting from the interaction $\pi + N \to K + \Lambda$ ($K^* + \Lambda$,
...). One has [21]
$${dN^{\Lambda} \over d^4x} = <\sigma > \rho_{\pi}(x) \ \rho_p(x)
\eqno(3.1)$$

\noindent where $\rho (x)$ are space-time dependent particle densities and $<
\sigma >$ is
the interaction cross-section - properly averaged over the momentum
distributions of the
colliding particles. We take $<\sigma > = 1.5$ mb which is the value of $
\pi^- + p \to K^0 +
\Lambda/\Sigma^0$ at its maximum near threshold. Beyond the threshold these
two-body
cross-sections decrease rapidly but quasi two-body interactions convert this
sharp decrease
into a mild increase. \par

We use cylindrical space-time variables~: the longitudinal proper time is $
\tau = \sqrt{t^2
- z^2}$ and the space-time rapidity $y = {1 \over 2} \ell n \left [ {t + z
\over t - z}
\right ]$. In these variables $d^4x = \tau \ d\tau \ dy\ d^2s$, where $
\vec{s}$ is the
transverse coordinate and $d^2s$ an element of transverse area. As
customary, we assume boost
invariance - i.e. that the space-densities $\rho_i(x)$ are independent of
$y$ -
and assume, furthermore, that the dilution in time of the density is mainly
due to
longitudinal motion. With this two assumptions we have [21]~:
$$\rho_i(\tau ,y, \vec{s}) = \rho_i(\tau , \vec{s}) {\tau_0 \over \tau} \ \
\ . \eqno(3.2)$$

\noindent Using (3.2) it is trivial to integrate Eq. (3.1) on the variable $
\tau$. One gets
$${dN_{\Lambda} \over dy} = \int d^2s {dN_{\pi} \over dy \ d^2s} {dN_p \over
dy \ d^2s}
\ 3<\sigma> \ell n \left [ (\tau + \tau_0)/\tau_0 \right ] \eqno(3.3)$$

\noindent where $\tau_0$ is the formation proper time and $\tau$ the time
during which the
final interaction takes place. We take $\tau_0$ = 1 fm and $\tau$ = 3 fm -
based on
interferometry measurements which indicate a very short time of particle
emission and a
freeze out time $\tau_0 + \tau \sim 4$ fm [23]. The factor 3 results from
the product of three
pion times two nucleon species, divided by a factor 2 - which is due to the
fact that one
out of two $\pi N$ combinations have a negligeably small cross-section. For
instance among
the $\pi^{\pm}p$ combinations only $\pi^-p$ has to be considered. \par

Let us now explain how to compute $dN_i/dy \ d^2s$ in Eq. (3.3). The
formulae given in
Section 2 (see for instance Eq. (2.3)) give the expression of $dN/dy$ -
which is the result of
integrating $dN/dy \ d^2s$ over $d^2s$. Actually, the only dependence on $
\vec{s}$ is
in the average values $\bar{n}_A$, $\bar{n}_B$ and $\bar{n}$. In the Glauber
model,
these average quantities are known as a function of $\vec{b}$ and $\vec{s}$.
In the
optical approximation one has
$$\bar{n}(\vec{b}, \vec{s}) = {AB \sigma \over \sigma_{AB}} T_A(\vec{b} -
\vec{s})
\ T_B (\vec{s}) \ \ \ , \eqno(3.4)$$
\noindent and
$$\bar{n}_A(\vec{b}, \vec{s}) = {A \over \sigma_{AB}} T_A(\vec{b} - \vec{s})
\ \sigma_{NA}(\vec{s}) \eqno(3.5)$$

\noindent where
$$\sigma_{NA}(\vec{s}) = 1 - \left ( 1 - \sigma \ T_A (\vec{s}) \right )^A \
\ \ .$$

\noindent Here $\sigma$ is the nucleon-nucleon inelastic cross-section and
$T_A(\vec{b})$
the nuclear profile - for which we use the standard Saxon-Woods expression.
Clearly,
when computing the r.h.s. of Eq. (3.3) one has to perform the products
$\bar{n}^2(\vec{b}, \vec{s})$, $\bar{n}_A^2(\vec{b}, \vec{s})$, $
\bar{n}_A(b, \vec{s})$
$\bar{n}(b, \vec{s})$ and integrate them over both $d^2s$ and $d^2b$. It is
easy to see
that such integrals of products are smaller than the product of integrals
$dN_{\pi}/dy$ times $dN_p/dy$ and that the ratio of the former over the
latter is
roughly proportional to $R_A^{-2} \sim A^{-2/3}$. Therefore, the number of $
\Lambda$'s
resulting from the final state interaction, will increase quite fast with
increasing
$A$. Indeed, at mid-rapidities, the product $dN_{\pi}/dy$ times $dN_p/dy$
increases with $A$
approximately as $A^{8/3}$ and thus the number of $\Lambda$'s increases
approximately as
$A^2$.  \par

In order to compute the rapidity distributions of $\Lambda$ resulting from
the final state
interaction (3.1) one has to compute first the densities of pions and
protons. This has been
done using the formulae in Appendix 1. \par

The rapidity distributions of $h^-$ in $NN$, $SS$ and $Pb$-$Pb$ collisions
are given in
Figs.~4 and 5. The agreement with experiment is reasonable. Note that, due
to the two
string approximation in $NN$, the rapidity distribution tends to be too
broad. Indeed, the
contribution of the $q_s$-$\bar{q}_s$ strings, although small at 200 GeV/c,
is concentrated
at mid-rapidities. This is partly compensated by choosing steeper diquark
fragmentation
functions (see Appendix 1). In any case, our aim here is not to get the best
possible
description of $h^-$ distributions, but only to have a reasonable form of
$dN_{\pi^-}/dy$
in order to use it in Eq.~(3.1). Following Ref.~4, we have taken $dN_{
\pi^-}/dy = 0.93$
$dN_{h^-}/dy$. \par

The Feynman $x$ distributions of proton and antiprotons in $pp$ collisions
are given in
Fig.~6. The agreement with data is quite good. Note that diffractive
dissociation, which
produces a peak near $x = 1$ in the $p$ distribution has not been included
in the model. The
corresponding results for peripheral and central $SS$ collisions are given
in Fig.~7.
Actually, both the data and the theoretical curves correspond to the
difference $p -
\bar{p}$. The shape of the rapidity distribution is well reproduced in the
model for
peripheral collisions but not for central ones. In the latter case, the
shape of the
experimental distribution is much flatter than the theoretical one. Note,
however, that the
average multiplicity is the same for the two distributions. This
disagreement between an
independent string model prediction and experiment is very significant and
we are going to
discuss it in some detail. First of all, in DPM the stopping power is
entirely controlled by
the momentum distribution functions. More precisely, when the average number
of collisions per
nucleon increases (i.e. when $\bar{\mu}_A$ in Eq.~(2.3) increases), the
maximum of the proton
spectrum is shifted to smaller rapidities (see Fig.~2 of Ref. [24]). This
effect is seen in
Fig.~7. However, in central $SS$ collisions $\bar{\mu}_A \sim 2$ and,
therefore, the effect is
small. Moreover, the decrease of the proton spectrum between its maximum and
$y^* \sim 0$ is
very steep even for a large number of inelastic collisions. Thus, the
dramatic flattening of
the proton rapidity distribution between peripheral and central $SS$
collisions in Fig.~7
cannot be reproduced in the model, and, as discussed in the Introduction,
rescattering of the
produced nucleons is needed. \par

We can now compute the rapidity distribution of $\bar{\Lambda}$ resulting
from the final state
interaction $\pi + \bar{N} \to K + \bar{\Lambda}$ ($K^* + \bar{\Lambda}$,
...).  The
results are shown in Fig.~2. The agreement with experiment is quite good.
\par

The corresponding results for $\Lambda$ production, including the effect of
the final
state interaction $\pi + N \to K + \Lambda$ ($K^* + \Lambda$, ...), are
given in Fig.~3 (dashed line). We see that the final state interaction has
produced a
substantial increase of the $\Lambda$ yield, which is now in qualitative
agreement with
experiment. We conclude that the main features of $\Lambda$ and $\bar{
\Lambda}$ production in
central $SS$ collisions can be understood in the scenario presented in this
paper. \par

We turn next to central $Pb$-$Pb$ collisions at 160 GeV/c per nucleon. In
this case the
rate of strange baryon production from the final state interactions is very
high at
mid-rapidities due to the large value of the pion density and one can not
ignore any longer
the crossed reactions $\pi + \Lambda \to K + N$ and $\pi + \bar{\Lambda }
\to K + \bar{N}$.
Indeed, it turns out that, at mid-rapidities, equilibrium between strange
and non strange
baryons is reached before the freeze out time $\tau_0 + \tau$ (assumed to be
the same as in
central $SS$ collisions). When this equilibrium is reached one has\break
\noindent $<n>_p +
<n>_n = 1.6 <n>_{\Lambda}$ (where $0.6 <n>_{\Lambda}$ takes into account the
$\Sigma^{\pm}$
baryons) and the strange baryon multiplicity does not increase any longer.
The predictions
for the rapidity distribution of $\bar{\Lambda}$ and $\Lambda$ in central
$Pb$-$Pb$ are shown
in Figs.~2 and 3 (upper lines scaled down by a factor~5). \par

A close look at Fig.~3 shows that the computed $\Lambda$ distribution in
$SS$ collisions
has a shallow dip at $y^* = 0$ which is not seen in the data. Probably this
is related to the
lack of baryon stopping discussed above and requires further rescattering of
the produced
$\Lambda$'s. In order to have a rough estimate of the effect of the baryon
rescattering, we
have introduced some extra stopping in the proton spectrum  (without
changing the proton
average multiplicity) by shifting it by $\Delta y = 0.5$. The modified $p-
\bar{p}$ rapidity
distribution is shown in Fig.~7 (full upper line). Using such shifted proton
distributions in
Eq. (3.3), we obtain a flatter $\Lambda$ distribution both in $SS$ and
$Pb$-$Pb$ collisions
(full lines in Fig.~3), which for $SS$ collisions is in good agreement with
experiment. \par

Finally, the average multiplicities of pions, kaons, protons lambdas and anti
\-lambdas in
$SS$ and $Pb$-$Pb$ collisions are given in Table~1. (A discussion of
cascades and omega
production in the present framework can be found in Refs.~10, 11). We
observe that the
ratio $\bar{\Lambda}/h^-$ ($\Lambda /h^-$) increase by about 50 $\%$ (70 $
\%$) between $SS$
and $Pb$-$Pb$ collisions.  \par \medskip

\noindent {\bf 4. Conclusions} \medskip
We have studied meson and baryon strangeness enhancement in an independent
string model with
quark-antiquark and diquark-antidiquark pairs in the nucleon sea. The ex\-pe
\-ri\-mental data
have forced us to depart from string independence and to introduce a final
state interaction
of co-moving secondaries. The necessity of final state rescattering also
follows from the
study of the shape of the proton rapidity distribution in central $SS$
collisions. Final state
interactions are the simplest forms of string interaction - which is a first
step in order
to drive the system toward equilibrium. Using all these ingredients, we have
obtained a
reasonable agreement with experiment. We feel that a more accurate
description is not
possible in the present state of the art. \par

The present model has many uncertainties that we want to discuss briefly.
Apart from the
poor knowledge of the parameter $S$ and $\alpha$, there is an uncertainty
resulting from the
hadronization of short strings and, in particular, of strings involving sea
diquarks (see
Appendix 1). Furthermore, the introduction of final state interactions rises
a lot of
questions. Not only a rigourous treatment of the final state interaction
does not exist,
but, moreover, we can not be sure that other reactions (for most of which
the cross-sections
are not even known) do not affect our results. (In particular we have
neglected
annihilation reactions of nucleons and antilambdas or antinucleons and
lambdas which
have a very large cross-section at threshold. However, this cross-section
drops very
fast after the threshold and the corresponding value of $<\sigma >$ in Eqs.
(3.1),
(3.3) is not known). Despite all these drawbacks, the effects discussed in
the present work
are important and cannot be ignored in strangeness enhancement studies. \par

A characteristic feature of the model presented here is that it predicts a
continuous
strangeness enhancement, not only from proton-proton to proton-nucleus and
to average (i.e.
minimum-bias) nucleus-nucleus collisions, but also in proton-proton when
energy or event
multiplicity is increased. Of course to get this latter effect it is
necessary to give up
the two string approximation in $pp$ collisions. This assumption is of no
consequence for the
results presented here, provided a good description of $pp$ data is achieved
[25].
Indeed the results for nucleus-nucleus collisions are determined by the
input in $pp$ (or
rather $NN$) collisions. However, when multistrings are introduced, one
obtains [26] an
increase of the ratio $K/\pi$ in $pp$ collision when increasing either the
energy or the
event multiplicity, as observed experimentally. Moreover, the increase with
energy
[7] of the ratio $\bar{\Lambda}/\bar{p}$, can also be understood in this
way. Note, however,
that the contribution to strangeness enhancement resulting from the final
state interaction
(which is very important for $\Lambda$ production) is most effective in
central
nucleus-nucleus collisions where the densities of secondaries are very
large. \par

In conclusion, the study of strangeness enhancement is a very active field.
This is clear
from the large number of recent work on the subject [1]. Actually, during
the latest
stages of the present work [27] many interesting papers on this subject have
appeared in
the literature [28-31]. Some of them give also predictions for $Pb$-$Pb$
collisions. The
theoretical ideas in these papers and their predictions are, in general,
quite different
from ours. Forthcoming results with a lead beam at CERN will certainly
trigger more
activity and give new insight into this very interesting field. Predictions
for RHIC and LHC
energies are also needed. \par

\vskip 2 mm
\noindent {\bf Acknowledgements :} \par
We would like to thank A. Krzywicki for useful discussions. We also thank J.
Ranft for an
ongoing collaboration on this subject. \par

One of the authors (C. M.) has benefitted of a EEC
postdoctoral project ERBCHBI CT 930 547. \par

This work has been done with the help of an INTAS contract 93-0079.

\vfill\supereject
\centerline{\bf Appendix 1} \bigskip
\noindent \underbar{\bf Hadron Spectra of Individual String} \par \medskip
Both in DPM and QGSM the hadron spectra of the individual strings are given
by a convolution
of structure functions and fragmentation functions. In DPM, one has for the
rapidity density
$N^{qq-q}(y, s)$ of hadron $h$ in, say, a diquark-quark string~:
$$N_{\mu_1,\mu_2}^{qq-q}(y,s) = \int_0^1 dx_1 \int_0^1 dx_2 \ \rho_{
\mu_1}^{qq}(x_1)
\ \rho_{\mu_2}^q(x_2)$$
$${dN^{qq-q} \over dy} (y - \Delta , s_{str.}) \ \theta (s_{str.} -
s_{thr.}) \ \
\ . \eqno(A.1)$$

\noindent Here $\rho_{\mu_1}^{qq}$ and $\rho_{\mu_2}^q$ are the momentum
distribution
functions of diquark and quark which depend on the light cone momentum
fractions $x_1$ and
$x_2$. The index $\mu_1$ $(\mu_2)$ denotes the number of inelastic
collisions suffered by
the nucleon to which the diquark (quark) belongs. The squared invariant mass
of the string
is $s_{str} = x_1x_2s$ and $\Delta = 1/2 \log [x_1/x_2]$ is the rapidity
distance between
the C. of M. of the string and the overall C. of M. At fixed values of $x_1$
and $x_2$ the
string hadronization is just given in terms of $qq$ and $q$ fragmentation
functions~:
$${dN^{qq-q} \over dy} (y - \Delta , s_{str.}) = \cases{
G_{qq}^h(y) \quad \hbox{for} \ y \geq \Delta \cr
\cr
G_q^h(y) \quad \ \hbox{for} \  y \leq \Delta \cr
} \eqno(A.2)$$
\noindent where $G(y)$ are fragmentation functions (see below). Of course
$dN^{qq_v-q_v}/dy
= 0$ for $y$ outside the rapidity interval determined by the string
endpoints. Note the
presence of a threshold $\theta$-function in Eq. (A.2). The value of
$s_{thr}$ corresponds
to the squared mass of the lightest system that can be produced in
association with the
triggered hadron $h$\footnote{*}{For instance for a $qq_v$-$q_v$ string
fragmenting into a
$\Lambda$ we have $s_{thr} = (m_{\Lambda} + m_K)^2$. For that of a $us$-$d$
string we have
$s_{thr} = m_{\Lambda}^2$. Likewise for pion (kaon) production in a $u
\bar{d}$ ($s\bar{d}$)
string we have $s_{thr} = \mu_{\pi}^2(m_K^2)$ - where $\mu_{\pi}$ is the
pion transverse
mass.}. \par

An important feature of DPM and QGSM is the fact that the structure function
is determined
in terms of Regge intercepts. In DPM, the structure function in the presence
of $2k$ such
constituents is given by [9]~:
$$\rho_k(x_1 \cdots x_{2k}) = C_{2k} \ x_1^{-1/2} x_2^{-1} \cdots
x_{2k-1}^{-1} x_{2k}^{3/2}
\delta \left ( 1 - \sum_{i=1}^{2k} x_i \right ) \ \ \ . \eqno(A.3)$$

\noindent The index 1 refers to the valence quark and the index $2k$ to the
diquark. All
other indices refer to sea quarks and antiquarks. The constant $C_{2k}$ is
obtained from the
normalization of $\rho_{2k}$ to unity. The structure function of each
individual
constituent, which appears in Eq. (A.1), is obtained by integrating (A.3)
over the
variables of all other constituents. (In actual calculations the singular
factor $x^{-1}$ is
replaced by $\bar{x}^{-1}$ with $\bar{x} = (x^2 + \mu^2/s)^{1/2}$, and $\mu$
= 0.1 GeV~;
varying $\mu$ between 0.05 and 0.3 produces only minor changes in our
results). \par

In QGSM the formulae are somewhat different [7]. The diquark and quark
fragment
independently from each other and the threshold resulting from the $
\theta$-function in
(A.1) is not present. Moreover in Eq. (A.3) sea quarks are assumed to have,
at present
energies, a $x^{-1/2}$ behaviour at $x \sim 0$ - the same as for valence
quarks. In this
case the integration of $\rho$ in (A.3) over all variables except one can be
performed
analytically. One gets
$$\eqalignno{
&\rho_{2k}^q(x) = C_{2k}^q \ x^{-0.5} (1 - x)^{2k + 1/2} \cr
&\rho_{2k}^{qq}(x) = C_{2k}^{qq} \ x^{1.5} (1 - x)^{2k- 3/2} \ \ \ . &(A.3')
\cr
}$$

\noindent The coefficients $C$ are determined (analytically) from the
normalization to unity
of $\rho (x)$. Note that the $\bar{x}^{-1}$ behaviour in DPM strongly
reduces the probability
of a sea quark to cross to the opposite hemisphere (as compared to an
$x^{-1}$ behaviour,
which corresponds to a flat distribution in rapidity). As a consequence, the
results
obtained in DPM and QGSM are quite similar - at least at the energies
considered here. \par

In the first DPM papers, fragmentation functions were just derived from a
fit to $pp$ (or $e^+e^-$) data. An important break through was achieved when
it was
realized that, not only momentum distribution functions, but also
fragmentation functions
could be derived in terms of Regge intercepts [7]. In this way the
predictivity of the
model was greatly increased. Indeed, it has been possible to predict the
shape of the
rapidity distributions of $\pi$, $K$, $p$, $\Lambda$, etc in $pp$ collisions
- with
absolute rates as the only free parameters. \par

In the following, we use the fragmentation functions of baryons and
antibaryons obtained in
this way. More precisely, the form of the fragmentation functions for $
\Lambda$ and
$\bar{\Lambda}$ are given by [20] $$\eqalignno{
&D_u^{\bar{\Lambda}} = D_d^{\bar{\Lambda}} = {1 \over Z} (1 - Z)^{3.5} \cr
&D_u^{\Lambda} = D_d^{\Lambda} = {1 \over Z} (1 - Z)^{2.5} \cr
&D_{uu,2}^{\bar{\Lambda}} = D_{ud,2}^{\bar{\Lambda}} = D_{dd}^{\bar{
\Lambda}} = {1 \over Z}
(1 - Z)^{5.5} \cr
&D_{uu,2}^{\Lambda} = D_{dd,2}^{\Lambda} = D_{ud,2}^{\Lambda} = (1 -
Z)^{6.5} \ \ \ .&(A.4)
\cr }$$

\noindent Here $D_{qq,2}$ denotes the fragmentation of a diquark into a
non-leading baryon.
We have to add to it a (dominant) term, $D_{qq,1}$, corresponding to leading
baryon
production, given by $$D_{uu,1}^{\Lambda} = D_{dd,1}^{\Lambda} = {1 \over Z}
Z^{1.5} (1 -
Z)^{1.5} \quad ; \quad D_{ud,1}^{\Lambda} = {1 \over Z} Z^{1.5} (1 -
Z)^{0.5} \ \ \ .
\eqno(A.5)$$

\noindent Turning to $p$ and $\bar{p}$ production, we have [20]
$$\eqalignno{
&D_u^{\bar{p}} = D_d^{\bar{p}} = {1 \over Z} (1 - Z)^3 \cr
&D_u^p = {1 \over Z} (1 - Z)^2 \quad ; \quad D_d^p = {1 \over Z} (1 - Z)^2
\left [ {1 \over
3} + {2 \over 3} (1 - Z) \right ] \cr
&D_{dd,2}^{\bar{p}} = D_{uu,2}^{\bar{p}} = D_{ud,2}^{\bar{p}} = {1 \over Z}
(1 - Z)^5 \cr
&D_{dd,2}^p = D_{uu,2}^p = D_{ud,2}^p = {1 \over Z} (1 - Z)^6  &(A.6) \cr
}$$
\noindent and, for the leading proton,
$$D_{ud,1}^p = {1 \over Z} Z^{1.5} (1 - Z)^0 \quad ; \quad D_{uu,1}^p = {1
\over Z} (1 - Z)^0
\ \ \ . \eqno(A.7)$$
\noindent Finally for neutron production we have
$$\eqalignno{
&D_u^n = D_d^p \quad ; \quad D_d^n = D_u^p \quad , \quad D_{uu,2}^n =
D_{uu,2}^p \ \ \ , \cr
&D_{ud,1(2)}^n = D_{ud,1(2)}^p \quad , D_{uu,1}^n = {2 \over 3Z} Z^{1.5} (1
- Z)^1 \ \
\ . &(A.8) \cr }$$

\noindent The values of all powers of $Z$ and $1 - Z$ in Eqs. (A.4)-(A.8)
are determined in
terms of Regge intercepts\footnote{*}{The only factors which are not
determined in this way
are the factor $[1/3 + 2/3 (1 - Z)]$ in Eq. (A.6) and the factor 2/3 in Eq.
(A.8). These
factors have a trivial origin. For instance the factor in (A.6) results from
the fact that
the fragmentation of $u$ and $d$ quarks into a proton is the same at $Z \to
0$. However, at
$Z \to 1$ the fragmentation of a $u$-quark is three times larger since the
three
combinations $u(ud)$, $u(du)$ and $u(uu)$ are possible. (The later will give
a
$\Delta^{++}$ which will often decay into a proton). On the contrary, for
the fragmentation
of a $d$ quark the only possible combination is $d(uu)$. Likewise, the
fragmentation of a
diquark $uu$ into a leading neutron is only possible through diquark
breaking and has a
probability of 2/3 as compared to the corresponding one for a $ud$
diquark.}. For
example, the power 0 of $1 - Z$ in Eq. (A.7) is given by $- \alpha_R(0) +
\lambda$ where
$\alpha_R(0) = 0.5$ is the dominant reggeon intercept and $\lambda = 2
\alpha 'p_{\bot}^2
\simeq 0.5$ is common to all fragmentation functions and results from the
integration over
transverse momentum. In the case of $\Lambda$ production $D_{ud,1}^{
\Lambda}$ has a softer $Z
\to 1$ behaviour, with a power $- \alpha_{\phi}(0) + \lambda = 0.5$, which
explains the
different shape of $\Lambda$ and $p$ (non-diffractive) rapidity
distributions (see Figs.~1
and 6). Of course, there are some numerical uncertainties in the
determination of these powers
- in particular in the power of $Z$ in the leading baryon fragmentation
function. The value
1.5 used above has been chosen, within the theoretically allowed range, in
order to obtain the
best description of the $pp$ data in the DPM formalism. \par

The only remaining point is to define the functions $G_{qq}$ and $G_q$ in
the r.h.s. of
Eq.~(A.2), in terms of the fragmentation functions defined above. This is
done in the
following way. Let us consider first the production of antibaryons and
non-leading baryons.
In this case we have $$\vbox{\eqalignno{
&G_{qq,2}^h(y) = \bar{a}_h \ Z_+ \ D_{qq,2}^h(Z_+) \ Z_- \ D_q^h(Z_-) \quad,
\qquad
\hbox{for} \ y \geq \Delta \cr
&G_{q,2}^h(y) = \bar{a}_h \ Z_- \ D_q^h(Z_-) \ Z_+ \ D_{qq,2}^h(Z_+) \quad ,
\qquad
\hbox{for} \ y \leq \Delta \ \ \ . &(A.9) \cr }}$$

\noindent Here $Z_+$ and $Z_-$ are the light cone momentum fractions defined
as $Z_+ = \exp
(y^{str} - y_{MAX}^{str})$, $Z_- = \exp (- y^{str} - y_{MAX}^{str})$, where
$y^{str} = y -
\Delta$ and $y_{MAX}^{str}$ is the maximal rapidity that a hadron $h$ can
have in the string
(for fixed values of the momentum fractions $x_1$ and $x_2$ of the string
ends). Note that
for a string with a very large invariant mass $Z_- \sim 0$ for $y \geq
\Delta$ and therefore
$G_{qq,2}^h(y) \sim \bar{a}_h \ Z_+ \ D_{qq}^h(Z_+)$. Likewise, for $y \leq
\Delta$, $Z_+
\sim 0$ and $G_{q,2}^h(y) \sim \bar{a}_h \ Z_- \ D_q^h(Z_-)$. \par

In the case of a leading baryon, Eq. (A.2) has to be modified. We have
$${dN^{qq-q} \over dy} \left ( y - \Delta , s_{str} \right ) = G_{qq,1}(y)
\eqno(A.2')$$

\noindent for both $y \geq \Delta$ and $y < \Delta$, with
$$G_{qq,1}^h(y) = a_h \ Z_+ \ D_{qq,1}^h(Z_+) \ \ \ , \eqno(A.10)$$

\noindent when the diquark is located in the positive rapidity hemisphere.
In the opposite
case one has to replace $Z_+$ by $Z_-$. Note that, when the string invariant
mass is large,
one has $Z_+ \ D_{qq,1}^h(Z_+) \sim 0$ in the backward hemisphere and,
therefore, the
leading baryon distribution is concentrated on the diquark side - as it
should be. A
further modification is necessary in order to preserve baryon number
conservation. Eqs.
(A.1) (A.2') do not satisfy this property - a main source of violation being
the threshold
factor. Actually, a non Monte Carlo formalism which leads to exact baryon
conservation has not
been found so far. However, a good numerical accuracy in this conservation
law is achieved by
dividing the r.h.s. of Eq. (A.1) by $$D = \int_0^1 dx_1 \int_0^1 dx_2 \
\rho(x_1) \ \rho(x_2)
\ \theta \left ( s_{str} - s_{thr} \right )  \ \ \ . \eqno(A.11)$$

\noindent This device to enforce baryon number conservation must be used
only for leading
baryon production. \par

The normalization factors $a_h$ and $\bar{a}_h$ have been determined in
order to reproduce
the measured rapidity distributions of baryons and antibaryons. We have~:
$$a_p = 0.82 \ \ , \quad \bar{a}_p = 0.036 \ \  , \quad a_{\Lambda} = 0.17 \
\  , \quad
a_{\bar{\Lambda}} = 0.012 \ \ \ . \eqno(A.12)$$

\noindent Note that the same normalization constant $\bar{a}$ is used for
antibaryons and
for non-leading baryons. \par

Plugging the above expressions for the rapidity distributions of the
individual strings in
Eq. (2.3) we can compute the baryon and antibaryon rapidity distributions in
nucleus-nucleus
collisions. The only missing ingredients are the fragmentation functions of
the strange sea
diquarks $us$, $ds$ and $ss$. For these functions we use the following
approximate relations
$$D_{us,1}^{\Lambda} = D_{ds,1}^{\Lambda} = D_{\bar{u}\bar{s},1}^{\bar{
\Lambda}} =
D_{\bar{d}\bar{s},1}^{\bar{\Lambda}} = D_{ud,1}^p \ \ \ ,$$ $$D_{ss,1}^{
\Lambda} =
D_{\bar{s}\bar{s},1}^{\bar{\Lambda}} = D_{uu,1}^n \ \ \ . \eqno(A.13)$$

\noindent Although everything is now fully especified in order to compute
baryon and
antibaryon rapidity distributions, we give, as an illustration, the detailed
form of the
fragmentation of a $qq_s$-$q_s$ string into a leading $\Lambda$ (which is,
of course,
identical to that of an $\bar{\Lambda}$ in a $\bar{q}\bar{q}_s$-$\bar{q}_s$
string). We have
$$N_{\bar{\mu}_A, \bar{\mu}_A}^{qq_s^P-q_s^T}(y) = {1 \over D} \int_0^1 dx_1
\int_0^1 dx_2
\ \rho_{\bar{\mu}_A}^{q_s}(x_1) \ \rho_{\bar{\mu}_A}^{q_s}(x_2) \ G_{qq,1}^{
\Lambda}(y) \
\theta \left ( s_{str} - s_{thr.} \right ) \eqno(A.14)$$

\noindent where
$$G_{qq,1}^{\Lambda}(y) = {4 \over (2 + S)^2} a_{\Lambda} Z_+ \left ( {2
\over 3}
D_{ud,1}^{\Lambda} (Z_+) + {1 \over 3} D_{uu,1}^{\Lambda}(Z_+) \right ) +$$
$${4S \over (2 + S)^2} a_p \ Z_+ \ D_{ud,1}^p(Z_+) + {S^2 \over (4 + S)^2}
a_p \ Z_+ \
D_{uu,1}^n(Z_+) \ \ \ . \eqno(A.15)$$

\noindent Eq. (2.4) is obtained from (A.1) after dividing by the denominator
in (A.11).
Note that the $\rho$ function of the sea diquark is taken to be the same as
the
corresponding one of a sea quark. In Eq. (2.5) we have explicitly written
the contributions
of the non-strange sea diquarks (terms proportional to $a_{\Lambda}$) and
that of the strange
sea diquarks (terms proportional to $a_p$). The contribution of the
non-strange sea
diquarks ($uu$, $dd$, $ud$ and $du$) is proportional to $4/(2 + S)^2$. That
of the diquarks
containing one strange quark ($us$, $ds$, $su$ and $sd$) is proportional to
$4S/(2 + S)^2$,
and that of the diquark $ss$ to $S^2/(2 + S)^2$. For the fragmentation
functions of the
strange sea diquarks we have used Eqs. (A.13). \par

Note that for the baryon production in strings containing sea diquarks we
have used the same
prescription as for valence diquarks, namely we have divided by the
denominator $D$ (see
Eqs. (A.11) and (A.14)) - although, in this case, this is not required by
baryon number
conservation since baryons are produced by pairs. Without dividing by $D$
the production of
$\Lambda$ and $\bar{\Lambda}$ from sea diquark strings would have been
substantially
smaller. Actually, we regard the hadronization of the sea diquark strings as
the main
source of uncertainty in this approach. \par

Turning next to the production of $p$ and $\bar{p}$, we have to replace in
Eq. (A.14),
$G_{qq,1}^{\Lambda}$ by
$$G_{qq,1}^p(y) = {4 \over (2 + S)^2} a_p \left [ {2 \over 3}
D_{ud,1}^p(Z_+) + {1 \over 6}
D_{uu,1}^p(Z_+) + {1 \over 6} D_{dd,1}^p(Z_+) \right ] \ \ \ . \eqno(A.16)$$

\noindent In this case, the contribution coming from the fragmentation of
strings containing
strange sea diquarks turns out to be very small and has been neglected. \par

In order to compute the final state interaction of secondaries (pions and
nucleons)
introduced in Section~3, we also need the rapidity distribution of $h^-$. In
this case the
contribution of the $q$-$\bar{q}$ strings is not negligeable and has the
effect of narrowing
the rapidity distribution. For our present purpose, however, it is enough to
consider the
two valence strings, provided the fragmentation function of valence quark
and diquark are
chosen in such a way as to reproduce the $NN \to h^-$ distribution. As
expected from
the absence of $q$-$\bar{q}$ strings, the diquark fragmentation functions
have to be softer
than expected from Regge arguments. (Of course this drawback disappears when
multistrings are
introduced [32]). We take $$G_{ud}^{\pi^-}(y) = a_{h^-} (1 - Z)^4 \ \ ;
\quad G_{uu}^{\pi^-}
= a_{h^-} (1 - Z)^5 \quad G_{dd}^{\pi^-} = a_{h^-} (1 - Z)^4 \eqno(A.17)$$

\noindent where
$$Z = \left | 2 \mu_{\pi} \sinh (y - \Delta)/\sqrt{s_{str}} \right | \ \
\ ,$$

\noindent and $\mu_{\pi}$ is the transverse pion mass. The normalization,
chosen to
reproduce the rapidity plateau height in $NN$ collisions, is $a_{h^-} =
0.59$. For the quark
fragmentation functions we take the ones introduced in Refs. [33], namely
$$\eqalignno{
&G_d^{h^-}(y) = {a_{h^-} \over 1.35} (1 + Z) \ F(Z) \cr
&G_d^{h^+}(y) = {a_{h^-} \over 1.35} (1 - Z) \ F(Z)  &(A.18) \cr
}$$

\noindent where
$$F(Z) = {1.3 (1 - Z)^2 + 0.05 \over 1 - 0.5Z} \ \ \ .$$

Actually, in computing the $h^-$ rapidity distribution in $AA$ collisions,
we have used Eq.
(2.2), i.e. the expression without sea diquarks. More precisely, we take
$${dN_{h^-}^{A_PA_T} \over dy} = \bar{n}_A \left ( N_{\bar{\mu}_A,
\bar{\mu}_A}^{qq_v^P-q_v^T}(y) + N_{\bar{\mu}_A, \bar{
\mu}_A}^{q_v^P-qq_v^T}(y) \right )
+$$
$$\left ( \bar{n} - \bar{n}_A \right ) {2 \over 2 + S} \left ( N_{\bar{
\mu}_A,
\bar{\mu}_A}^{q_s^P-\bar{q}_s^T}(y) + N_{\bar{\mu}_A, \bar{\mu}_A}^{
\bar{q}_s^P-q_s^T}(y)
\right )  \eqno(A.19)$$

\noindent In Eq. (A.19) we only consider the contribution of the non-strange
sea
strings. Of course some pions are also produced in strange sea ones. However
their
contribution is small and roughly compensates the loss of pions resulting
from the
introduction of sea diquark strings (see Section 2), where pion production
is smaller than
in $q_s$-$\bar{q}_s$ strings. \par

Finally, to compute the kaon multiplicities (Table 1), we have used the
experimental values
in $NN$ collisions [6] and have assumed 11] that the fragmentation $s \to
K^+$ ($\bar{s}
\to K^-$) is equal to $u \to \pi^+$ ($d \to \pi^-$). This is, of course,
justified only at
the first string break-up. However, due to the small momentum fraction
carried by the sea
quark the probability of further break-ups in the rapidity hemisphere of the
sea quark is
negligeably small.

  \vfill\supereject \centerline{\bf References}
\bigskip
 \item{[1]} See Proceedings of S'95 Conference, Tucson, January 1995.
\item{[2]} B. Koch, B. Muller and J. Rafelski, Phys. Rep. {\bf 142} (1986)
167.
\item{[3]} R. Salmeron, Nucl. Phys. {\bf A566} (1994) 199c.
\item{[4]} NA35 Collaboration~: T. Alber et al., Z. Phys. {\bf C64} (1994)
195.
\item{[5]} E755 Collaboration~: N. N. Biswas, Proceedings XXI International
Symposium on
Multiparticle Dynamics, World Scientific (1992).
\item{[6]} M. Gazdzicki and Ole Hansen, Nucl. Phys. {\bf A528} (1991) 754.
\item{[7]} E765 Collaboration~: T. Alexopoulos et al., Phys. Rev. {\bf D46}
(1992) 2773.
\item{[8]} NA36 Collaboration~: J. M. Nelson, Nucl. Phys. {\bf A566} (1994)
217c.
\item{[9]} A. Capella, U. Sukhatme, C. I. Tan and J. Tran Thanh Van, Phys.
Rep.
{\bf 236} (1994) 225.
\item{[10]} J. Ranft, A. Capella, J. Tran Thanh Van, Phys. Lett. {\bf B320}
(1994)
346~; \item{} H. J. M\"ohring, J. Ranft, A. Capella, J. Tran Thanh Van,
Phys. Rev.
{\bf D47} (1993) 4146 (the calculations in these papers are based on the
DPMJET and
DTNUC codes).  \item{[11]} A. Capella, Orsay preprint LPTHE 94-113.
\item{[12]} QGSM~: A. Kaidalov, Phys. Lett. {\bf B117} (1982) 459~; A.
Kaidalov and
K. A. Ter-Martirosyan, Phys. Lett. {\bf B117} (1982) 247. For the
corresponding Monte
Carlo code see N. S. Amelin et al., Phys. Rev. {\bf C47} (1993) 2299~; N. S.
Amelin et al,
Nucl. Phys. {\bf A544} (1992) 463c.  \item{[13]} A.D. Martin, R. G. Roberts
and W. J.
Stirling, preprint RAL-94-005 (DTP/94/3).
\item{[14]} N. Armesto, M.A. Braun, E.G. Ferreiro
and C. Pajares, University of Santiago de Compostela, preprint US-FT/16-94.
References to
earlier papers on string fusion can be found in G. Gustafson, Nucl. Phys. {
\bf A566} (1994)
233c. \item{[15]} RQMD~: H. Sorge, R. Matiello, A. von Kectz, H. St\"ocker
and W. Greiner, Z.
Phys. {\bf C47} (1990) 629~; H. Sorge, M. Berenguer, H. St\"ocker and W.
Greiner, Phys.
Lett. {\bf B289} (1992) 6~; Th. Sch\"onfeld et al, Nucl. Phys. {\bf A544}
(1992) 439c.
\item{[16]} FRITIOF~: B. Andersson, G. Gustafson and B. Nilsson-Almquist,
Nucl . Phys. {\bf
B 281} (1987) 289~; B. Nilsson-Almquist, E. Stenlund, Comp. Phys. Comm. {\bf
43} (1987)
387. In this code the enhancement of $\Lambda$ and $\bar{\Lambda}$ is
essentially due to
the final state interaction (B. Andersson, private communication).
\item{[17]} NA35 Collaboration~: H. Str\"obele et al., Nucl. Phys. {\bf
A525} (1991) 59c.
\item{[18]} JETSET~: T. Sj\"ostrand, CERN Report CERN-TH 6488/92, 1992.
\item{[19]} BAMJET~: S. Ritter, Comput. Phys. Commun. {\bf 31} (1984) 393.
\item{} J. Ranft and S. Ritter, Acta Phys. Pol. {\bf B11} (1980) 259.
\item{[20]} A. B. Kaidalov and O. I. Piskunova, Z. Phys. {\bf C30} (1986)
141.
\item{[21]} P. Koch, U. Heinz and J. Pitsut, Phys. Lett.
{\bf B243} (1990) 149.
\item{[22]} K. G. Boreskov, A. B. Kaidalov, S. M. Kiselev and N. Ya
Smorodinskaya, Sov. J.
Nucl. Phys. {\bf 53} (1991) 356.
\item{[23]} NA35 Collaboration~: G. Roland, Nucl. Phys. {\bf A566} (1994)
527c~; D. Ferenc, Proceedings XXIX Rencontres de Moriond (1994) ed. J. Tran
Thanh Van~;
\item{} NA44
collaboration~: T.J. Humanic, Nucl. Phys. {\bf A566} (1994) 115c~; S. Panday,
Proceedings Rencontres de Moriond, ibid.
\item{[24]} A. Capella, J. A. Casado, C. Pajares, A. V. Ramallo and J. Tran
Thanh Van, Z.
Phys. {\bf C33} (1987) 541.
\item{[25]} For Monte Carlo analysis of strangeness enhancement in the
present approach,
in\-clu\-ding multistrings in $pp$ collisions, see J. Ranft, A. Capella and
J. Tran Thnah Van,
in preparation.
\item{[26]} J. Ranft, preprint LNF 94/035 P.
\item{[27]} A short description of the results of the present work was given
in A. Capella et
al., Proceedings XXX Rencontres de Moriond, Les Arcs (France), 1995,
presented by C. Merino.
\item{[28]} T. S. Bir\'o, P. Levai and J. Zim\'anyi, preprint RIPNP,
Budapest (1995) and
Ref. 1.
\item{[29]} P. L\'evai and Xin-Nian Wang, preprint RIPNP, Budapest and Ref.
1.
\item{[30]} C. Slotta, J. Sollfrank and U. Heinz preprint TPR-95-04.
\item{[31]} K. Kadija, N. Schmitz and P. Seyboth, MPI-PhE/95-07.
\item{[32]} A. B. Kaidalov and O. I. Piskunova, Sov. J. Nucl. Phys. {\bf 41}
(1985) 816.
\item{[33]} A. Capella and J. Tran Thanh Van, Z. Phys. {\bf C10} (1981) 249.
\item{[34]} NA35 Collaboration~: D. R\"ohrich, Nucl. Phys. {\bf A566} (1994)
35c.
\item{[35]} NA35 Collaboration~: T. Alber et al., Proceedings XXX Rencontres
de Moriond,
Les Arcs, France (1995).
\item{[36]} A. E. Brenner et al., Phys. Rev. {\bf D26} (1982) 1497.

\vfill\supereject
\centerline{\bf Figure Captions} \bigskip
{\parindent=1 cm
\item{\bf Fig. 1} Feynman $x$ distributions of $\Lambda$ and $\bar{\Lambda}$
produced in $pp$
collisions at 200 GeV/c. The compilation of experimental data, at energies
ranging from
100 GeV/c to $\sqrt{s}$ = 63 GeV, is from Ref. 20. Error bars in the $
\Lambda$ data are
only shown for $x = 0.2$ and 0.95.  \vskip 5 mm

\item{\bf Fig. 2} Rapidity distribution of $\bar{\Lambda}$ in central $SS$
collisions at
200 GeV/c per nucleon (dashed line) compared to the data of Ref. 4. The
dashed-dotted
line is the corresponding result without final state interaction. The full
line is the
prediction for central $Pb$-$Pb$ at 160 GeV/c per nucleon, scaled down by a
factor 5.
\vskip 5 mm

\item{\bf Fig. 3} Rapidity distribution of $\Lambda$ in central $SS$
collisions at 200
GeV/c per nucleon (full and dashed lines) compared to the data of Ref. 4.
The dashed-dotted
line is the corresponding result without final state interaction. The dashed
(full) line is
obtained by adding the final state interaction, Eq. (3.3), with the proton
spectrum
corresponding to the dashed (full) lines in Fig.~7. The upper lines are the
predictions for
central $Pb$-$Pb$ collisions at 160 GeV/c per nucleon scaled down by a
factor 5. The
dashed (full) line is computed by adding the final state interaction with
the proton
spectrum computed in DPM (shifted by $\Delta y = 0.5$).  \vfill \supereject

\item{\bf Fig. 4} The rapidity distribution of negative hadrons ($h^-$) in
central $SS$
collisions (full line) and in $N$-$N$ collisions (dashed line), the latter
scaled up by a
factor 32, compared to the data of Ref. [34].  \vskip 5 mm

\item{\bf Fig. 5} Rapidity distribution of negative hadrons ($h^-$) in
central $Pb$-$Pb$
collisions at 160 GeV/c per nucleon. The data are from Ref. [35]. \vskip 5 mm

\item{\bf Fig. 6} Feynman $x$ distributions of proton and antiprotons in
$pp$ collisions at
200 GeV, compared with the data of Ref. [36] at 175 GeV/c. Diffraction is
not included in the
model calculation.  \vskip 5 mm

\item{\bf Fig. 7} Rapidity distribution of proton minus antiproton in
peripheral (full
line) and central (full and dashed lines) $SS$ collisions compared with data
from Ref. [17].
The theoretical curve for the peripheral case corresponds to $N + N \to (p -
\bar{p}) + X$,
normalized to the data. The dashed line in $SS$ is the result of DPM. The
full line in $SS$ is
obtained by adding some extra stopping to the proton (with $\Delta y = 0.5$)
without changing
the proton average multiplicity (see main text).  \vskip 5 mm

\vfill\supereject
$$\epsfbox {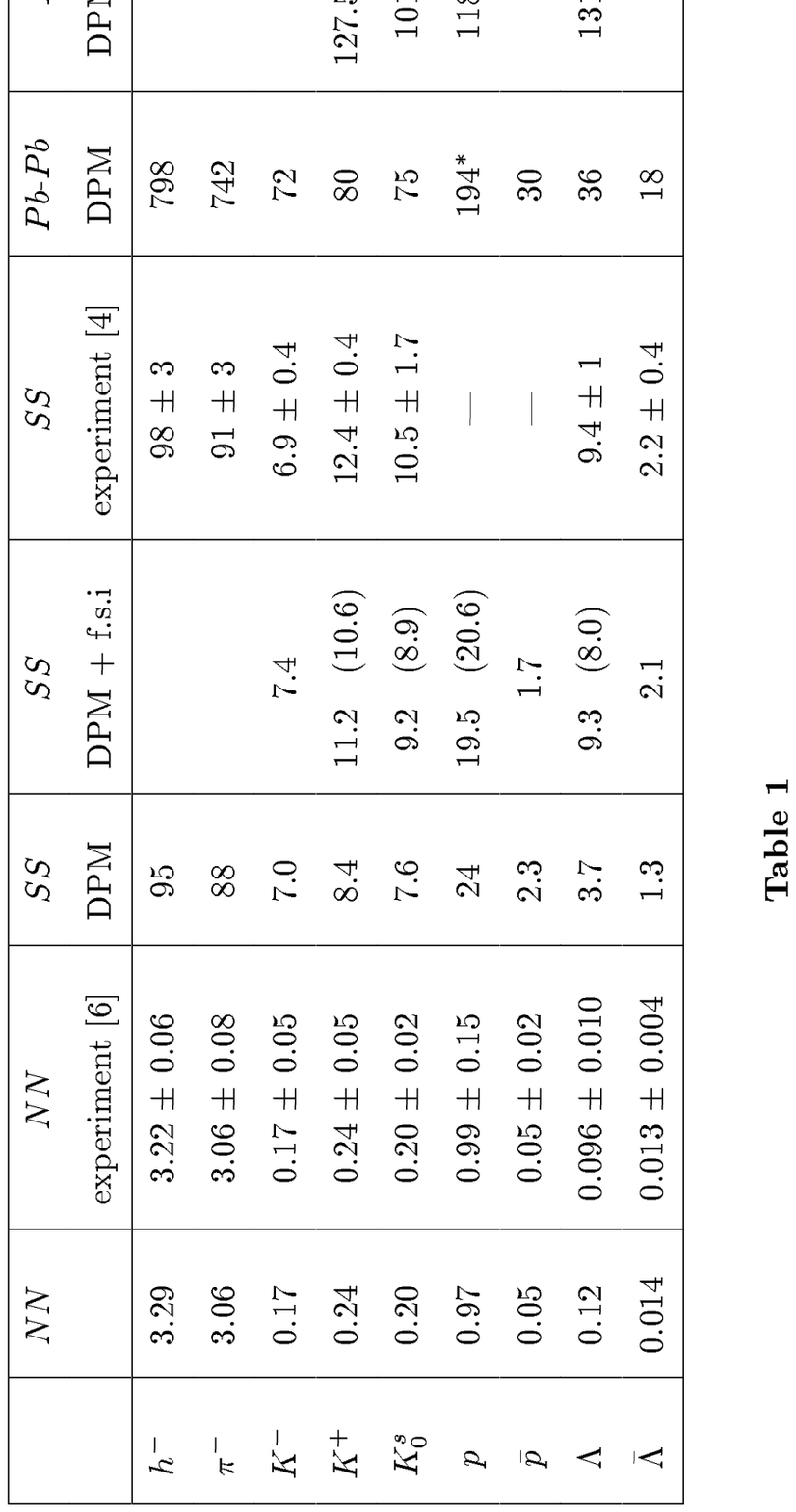}$$
\centerline{\bf Table Caption}
Average multiplicities of secondaries in $NN$, central $SS$ collisions at
200 GeV/c per
nucleon and central $Pb$-$Pb$ at 160 GeV/c per nucleon, compared with
available data. For
nucleus-nucleus collisions the multiplicities of $\Lambda$, $K^+$ and
$K_0^s$ in DPM are
also given before the final state interactions $\pi N \to K \Lambda$ and $
\pi \bar{N} \to
K \bar{\Lambda}$ are taken into account. The number of protons in $Pb$-$Pb$
(followed by an
asterix), corresponds to the combination $(p + n)/2$. The values of
multiplicities within
(without) a parenthesis correspond to the dashed (full) lines in Fig.~3. The
nucleon
multiplicities after final state interaction are obtained from the relation
$n_{p(\bar{p})}^{DPM + f.s.i} = n_{p(\bar{p})}^{DPM} - \left ( n_{\Lambda
(\bar{\Lambda)}}^{DPM + f.s.i} - n_{\Lambda (\bar{\Lambda})}^{DPM} \right )
\times 1.6/2$
(see main text).  \par

$$\epsfbox {fig1.p}$$
$$\epsfbox {fig2.p}$$
$$\epsfbox {fig3.p}$$
$$\epsfbox {fig4.p}$$
$$\epsfbox {fig5.p}$$
$$\epsfbox {fig6.p}$$
$$\epsfbox {fig7.p}$$

\bye